\documentclass[a4paper, 10pt, conference]{ieeeconf}
\IEEEoverridecommandlockouts
\usepackage{epsfig} 

\title{\LARGE \bf
Examples of Synchronization in Discrete Chaotic Systems}

\author{Juan C. Botero and Jean-Jacques E. Slotine
\thanks{This work was supported by the Fondazione Rocca Fellowship
through the MIT--Italy Program, Massachusetts Institute of
Technology and Politecnico di Milano.}
\thanks{Juan C. Botero is a PhD student in the Mechanical Systems
Engineering doctoral program at Politecnico di Milano (Technical
University). Dipartimento di Meccanica, Via La Masa 34, 20158,
Milan, Italy  {\tt\small juan.botero@ polimi.it}}%
\thanks{Jean--Jacques E.
Slotine is Professor of Mechanical Engineering \& Information
Sciences, Professor of Brain and Cognitive Sciences and Director of
the Nonlinear Systems Laboratory at Massachusetts Institute of
Technology, 77 Massachusetts Ave., Cambridge, MA 02139 USA
  {\tt\small jjs@mit.edu}}%
}

\begin{document}

\maketitle
\thispagestyle{empty}
\pagestyle{empty}

\begin{abstract}

This paper presents an application of partial contraction analysis
to the study of global synchronization in discrete chaotic systems.
Explicit sufficient conditions on the coupling strength of networks
of discrete oscillators are derived. Numerical examples and
applications to simple systems are presented. Previous researches
have shown numerically that the systems under study, when arranged
in a network, exhibits rich and complex patterns that can
dynamically change in response to variations in the environment.  We
show how this ``adaptation'' process strongly depends on the
coupling characteristics of the network. Other potential
applications of synchronized chaotic oscillators are discussed.

\end{abstract}

\section{INTRODUCTION}

This paper presents a research focused on coupled discrete dynamical
systems [1], [9]. In particular, it is of interest to study the
synchronization phenomenon in chaotic systems, i.e. systems in which
small differences on the initial conditions can lead to a completely
different behaviours in time. In spite of this characteristic,
synchronization can be achieved in this type of systems [10], [11].
This work presents a number of examples in which synchronization can
be guaranteed by choosing the appropriate coupling strength. The
coupling strength can be determinate using partial contraction
analysis. This analysis is able to predict complete synchronization
independently of the initial conditions, as long as the coupled
system verifies certain properties. Besides synchronization,
diffusively coupled dynamical systems oscillating chaotically in
time have seen to lead to interesting emerging properties due to
changes in the environment [$5$]. This feature has been used in this
work to simulate the motion of a two--legged robot where each leg is
a chaotic oscillator and these oscillators are diffusively coupled.
The appeal of such a system is the fast adaptation it shows when
sudden unexpected dynamic changes in the environment occur. This
method of locomotion does not require any control or optimization
process to be performed. Furthermore, this type of analysis can be
done on any periodical task that requires fast responses to
unpredictable changes.

The role of chaos in ``adaptation'' and evolution has been widely
discussed before [4], [8]. In particular, some research in this
area, sometimes known as ``adaptation at the edge of chaos'',
suggests the possibility that when biological systems adapt in order
to survive, the process of evolution may favor those systems that
are near a phase transition from order to chaos [8]. In any event,
the chaotic behaviour of simple dynamical systems can be exploited
to obtain adaptation behaviors to external dynamical disturbances.

The method used is mathematically simple and also similar to for the
analysis of coupled dynamical systems [3], [9], [10].  Defining
proper matrices to project the states onto the so-called
synchronization manifold [10], [11], contraction theory [7], [13],
[14], [17] can be used to determine conditions for which a set of
coupled identical systems will completely synchronize.

\section{Mathematical Foundations}

As in [$13$], consider a set of coupled dynamical systems and define
the vector {x\{\}}:

\begin{equation} \label{GrindEQ__1_} {{x}}_{\{ \} } =\left[\begin{array}{c} {x_{1} } \\ {\begin{array}{l} {x_{2} } \\ {{\rm \; }\vdots } \\ {x_{n} } \end{array}} \end{array}\right]\Rightarrow \dot{{x}}_{\{ \} } =\left[\begin{array}{c} {\dot{x}_{1} } \\ {\begin{array}{l} {\dot{x}_{2} } \\ {{\rm \; }\vdots } \\ {\dot{x}_{n} } \end{array}} \end{array}\right] \end{equation}

Define the matrices \textbf{U} and \textbf{V}

\begin{equation} \label{GrindEQ__2_} \textbf{U}=\left[\begin{array}{cccccc} {1} & {1} & {0} & {0} & {\cdots } & {0} \\ {0} & {1} & {1} & {0} & {\cdots } & {0} \\ {\vdots } & {\vdots } & {\vdots } & {\vdots } & {\cdots } & {\vdots } \\ {1} & {0} & {0} & {0} & {\cdots } & {1} \end{array}\right]\end{equation}
\begin{equation} \label{GrindEQ__3_} \textbf{V}=\left[\begin{array}{cccccc} {1} & {-1} & {0} & {0} & {\cdots } & {0} \\ {0} & {1} & {-1} & {0} & {\cdots } & {0} \\ {\vdots } & {\vdots } & {\vdots } & {\vdots } & {\cdots } & {\vdots } \\ {-1} & {0} & {0} & {0} & {\cdots } & {1} \end{array}\right] \end{equation}
\bigskip
such as

\begin{equation} \label{GrindEQ__4_} \begin{array}{l} {x^{||} =\textbf{U}x_{\{ \} } \Rightarrow \dot{x}^{||} =\textbf{U}\dot{x}_{\{ \} } } \\ {x^{\bot } =\textbf{V}x_{\{ \} } \Rightarrow \dot{x}^{\bot } =\textbf{V}\dot{x}_{\{ \} } } \end{array} \end{equation}

By construction

\begin{equation} \label{GrindEQ__5_} \textbf{UV}^{T} =\left[\begin{array}{cccccc} {0} & {1} & {0} & {0} & {\cdots } & {0} \\ {-1} & {0} & {1} & {0} & {\cdots } & {0} \\ {0} & {-1} & {0} & {1} & {\cdots } & {0} \\ {\vdots } & {\vdots } & {\vdots } & {\vdots } & {\cdots } & {\vdots } \\ {0} & {0} & {0} & {-1} & {\cdots } & {1} \\ {0} & {0} & {0} & {\cdots } & {-1} & {0} \end{array}\right] \end{equation}

\bigskip
so $<$ \textbf{U$_{i}$},\textbf{V$_{i}$} $>$  = 0, where
$<$\textit{x},\textit{y}$>$ denotes standard  inner
 product.
 Thus, the elements of the vectors \textbf{$x^{\|}$} and
\textbf{$x^{\bot}$} are orthogonal ``one to one'':
\bigskip

$\begin{array}{l} {{x}^{||} _{i} =\textbf{U}_{i} {x}_{i} ||\textbf{U}_{i} } \\
{{x}^{\bot } _{i} =\textbf{V}_{i} {x}_{i} ||\textbf{V}_{i} }
\end{array}$ $\rightarrow$ ${x}^{||} _{i} \bot {x}^{\bot } _{i} $
\bigskip
\section{Contraction Theory for Discrete Systems}

Some basic results in Contraction Theory are presented as follows:
\bigskip

\textit{\underbar{Definition 1}}

Given the systems of equations $x_{i+1} =f_{i} \left(x_{i}
,i\right)$ , a region of the state space is called
\textit{contraction region} with respect to a uniformly positive
definite metric  $M_{i} \left(x_{i} ,i\right)=\theta _{i} ^{T}
\theta _{i} $  if in that region $[7]$:

\begin{equation} \label{GrindEQ__6_} \exists \beta >0,F_{i} ^{T} F_{i} -I\le -\beta I<0 \end{equation}

where

\begin{equation} \label{GrindEQ__7_} F=\theta _{i+1} \frac{\partial f}{\partial x} \theta _{i+1} ^{-1}  \end{equation}

\bigskip

\textit{\underbar{}}\textit{\underbar{Theorem 1}}

Given the systems of equations  $x_{i+1} =f_{i} \left(x_{i}
,i\right)$ , any trajectory, which starts in a ball of constant
radius with respect to the metric $M_{i} \left(x_{i} ,i\right)$,
centered about a given trajectory and contained at all times in a
contraction region with respect to the metric $M_{i} \left(x_{i}
,i\right)$, remains in that ball and converges exponentially to this
trajectory. Furthermore, global exponential convergence to the given
trajectory is guaranteed if the whole state space is a contraction
region with respect to the metric $M_{i} \left(x_{i} ,i\right)$.
This corresponds to a necessary and sufficient condition for
exponential convergence of the system $[7]$.

\bigskip

\textit{\underbar{Theorem 2}}

Consider two coupled systems. If the dynamics equations verify

\begin{equation} \label{GrindEQ__8_} x^{\left(A\right)} _{n+1} -h\left(x^{\left(A\right)} _{n} \right)=x^{\left(B\right)} _{n+1} -h\left(x^{\left(B\right)} _{n} \right) \end{equation}

where the function \textit{h} is contracting, then $x^{(A)}$ and
$x^{(B)}$ will converge to each other exponentially, regardless of
the initial conditions $[17]$.

Hence, the process of one variable converging exponentially to
another, i.e. synchronization, can be seen equivalently as
contraction of the projection in the transverse manifold [$13$]. It
is clear since the transverse manifold is, by construction,
perpendicular to the synchronization manifold:

\begin{equation} \label{GrindEQ__9_} x_{1} \to x_{2} \Leftrightarrow x^{\bot } \to 0 \end{equation}

If the projection of a particular solution has null norm in the
transverse manifold, it lays entirely on the synchronization
manifold (here there are no distinction between synchronization and
oscillator dead). Thus, it is only necessary to study the hereafter
called error dynamics \textbf{$x_{i} - x_{j}$}, and prove its
contracting behaviour. The proof can be seen in $[13]$ and it can be
express in rather intuitive terms: synchronization will be achieved
if the difference of the states of any two oscillators is zero.

\bigskip
\section{Application of Contraction to Chaotic Systems Synchronization}

\subsection{General Case: Chaotic Maps}

In order to study the phenomenon of complete synchronization in
chaotic systems, it is usual to work with one-dimensional maps [5],
[12] due to the complex behaviours that can be observed from such
simple models. When no coupling is present between any of the maps,
due to their chaotic nature the systems behave random-like. But when
appropriate coupling is build between the maps, one will expect some
sort of interaction. Hence, in the case of synchronization or
complete synchronization, it is of interest to have a coupling with
two main properties [12]:

\begin{enumerate}
\item The coupling must make the states of the systems closer to each other, i.e. dissipative coupling.

\item The coupling must not affect the synchronization state.
\end{enumerate}

A general form of this kind of coupling operator of dimension-2 is:

\begin{equation} \label{GrindEQ__10_} L=\left[\begin{array}{cc} {1-\alpha } & {\beta } \\ {\beta } & {1-\beta } \end{array}\right] \end{equation}

\bigskip

where 0  $<$  \textit{a}  $<$  1 and 0  $<$  \textit{b}  $<$  1. The
simplest case of this coupling would be $\alpha$ = $\beta$ =
$\varepsilon$ [12]. This case allows the coupling to be symmetric,
which leads to a great simplification in the computations needed,
and furthermore, it allows the direct application of the proposed
method. Hence, it is possible to couple any two maps \textit{x} and
\textit{y} as [12]:

\begin{equation} \label{GrindEQ__11_} \left[\begin{array}{c} {x_{t+1} } \\ {y_{t+1} } \end{array}\right]=\left[\begin{array}{cc} {1-\varepsilon } & {\varepsilon } \\ {\varepsilon } & {1-\varepsilon } \end{array}\right]\left[\begin{array}{c} {f\left(x_{t} \right)} \\ {f\left(y_{t} \right)} \end{array}\right] \end{equation}
\bigskip

and with a proper choice of the coupling strength $\varepsilon$, the
complete synchronization state can be achieved. For example, two
oscillators following the well known skew map [8], [12]:

\[f\left(x_{n} \right)=\left\{\begin{array}{l} {\frac{x_{n} }{a} {\rm \; \; \; \; if\; \; \; }0\le x_{n} \le a} \\ {\frac{\left(1-x_{n} \right)}{\left(1-a\right)} {\rm \; \; if\; \; \; }a\le x_{n} \le 1} \end{array}\right. \]

which exhibits pure chaotic behaviour due to the subsequent
stretching and folding processes in the interval from 0 to 1 [12].
Two discrete systems of this type, say \emph{x} and \emph{y}, can be
coupled as in ($11$). Thus, varying the coupling strength
$\varepsilon$ it is possible to achieve complete synchronization of
the states. The complete proof of this example can be found in [12],
where the well known technique of \textit{transverse Lyapunov
exponents} is used [10], [11].

However, theorem 2 from contraction theory provides a simple yet
powerful tool to analyze this particular system. The following
theorem was presented for the first time in [17]. Using this theorem
it is possible to write the system of two skew maps \textit{x} and
\textit{y}, coupled together in the following form:

\begin{equation} \label{GrindEQ__12_} x_{n+1} -\left(1-2\varepsilon \right)f\left(x_{n} \right)=y_{n+1} -\left(1-2\varepsilon \right)f\left(y_{n} \right) \end{equation}

Hence, it is only necessary to verify that the function

\begin{equation} \label{GrindEQ__13_} h\left(x\right)=\left(1-2\varepsilon \right)f\left(x\right) \end{equation}

\bigskip

is contracting for \textit{f(x)} in ($12$) (for a particular
metric). As \textit{$\partial$f(x)/$\partial$x} is not continuous,
two cases must be taken into account. Each one of these cases should
give an interval for \textit{e} in which contraction of the function
\textit{h(x)} is guaranteed.\textit{ }The general result is the
intersection of these two intervals (in fact, the \textit{supremum}
or the biggest set contained in the intersection of the intervals
obtained). In this particular example, we must show that the
Jacobian

\begin{equation} \label{GrindEQ__14_} F=\frac{\partial h\left(x\right)}{\partial x} =\left(1-2\varepsilon \right)\frac{\partial f\left(x\right)}{\partial x}  \end{equation}

\bigskip

is contracting in some metric \textit{M}. Indeed, if \textit{M} is
equal the identity matrix \textit{I}, for the two cases we find the
suitable intervals

\[\begin{array}{l} {0\le x\le a{\rm \; \; \; }\to {\rm \; \; }\frac{\partial f\left(x\right)}{\partial x} ={\rm \; \; \; }\frac{1}{a} {\rm \; \; \; \; ;\; \; \;} {\rm \; \; \; \; }\frac{a-1}{2} <{\rm \; \; }\varepsilon {\rm \; \; }<\frac{a+1}{2} }\\ \\{a\le x\le 1{\rm \; \; \; }\to {\rm \; \; }\frac{\partial f\left(x\right)}{\partial x} =\frac{1}{a-1} {\rm \; \; ;\; \; }{\rm \; \; \; \; \; \; \; }\frac{a}{2} {\rm \; \; \; }<{\rm \; \; }\varepsilon {\rm \; \; }<1-\frac{a}{2} } \end{array}\]

\bigskip
Thus, the supremum of the intersection of the two intervals is the
second one (as it is contained in the first interval). It is
straight forward to verify that the use of the transverse manifold
leads to the exactly same conditions as before, i.e. an interval for
$\varepsilon$ in which the error dynamics is contracting \textit{in
the same metric} as before. In fact, the two synchronization
intervals obtained for the error are

\[\begin{array}{l} {0\le x\le a{\rm \; \; \; }\to {\rm \; \; }e{}_{n+1} =\frac{\left(1-2\varepsilon \right)}{a} e{\rm \; \; ;\; \; }\frac{a-1}{2} <{\rm \; \; }\varepsilon {\rm \; \; }<\frac{a+1}{2} } \\\\ {a\le x\le 1{\rm \; \; \; }\to {\rm \; \; }e{}_{n+1} =\frac{\left(1-2\varepsilon \right)}{a-1} e{\rm \; \; \; ;\; \; }\frac{a}{2} {\rm \; \; \; }<{\rm \; \; }\varepsilon {\rm \; \; }<1-\frac{a}{2} } \end{array}\]

\bigskip
Going beyond the analysis done by transverse Lyapunov exponents, the
contraction theory is capable of giving intervals where the selected
coupling strength $\varepsilon$ guarantees complete synchronization.
This can be clearly seen in figure 1. For two coupled oscillators
\textit{X} and \textit{Y}  and \textit{a} = 0.7, the proposed method
predicts complete synchronization (in which the values lie on the
\textit{X} = \textit{Y} line) for a coupling strength between 0.35
and 0.65. Fig. 1 shows two cases.
\begin{figure}[thpb]
      \centering
      \includegraphics[bb=30mm 30mm 208mm 296mm, width=40.9mm, height=33.1mm, viewport=3mm 4mm 205mm 292mm]{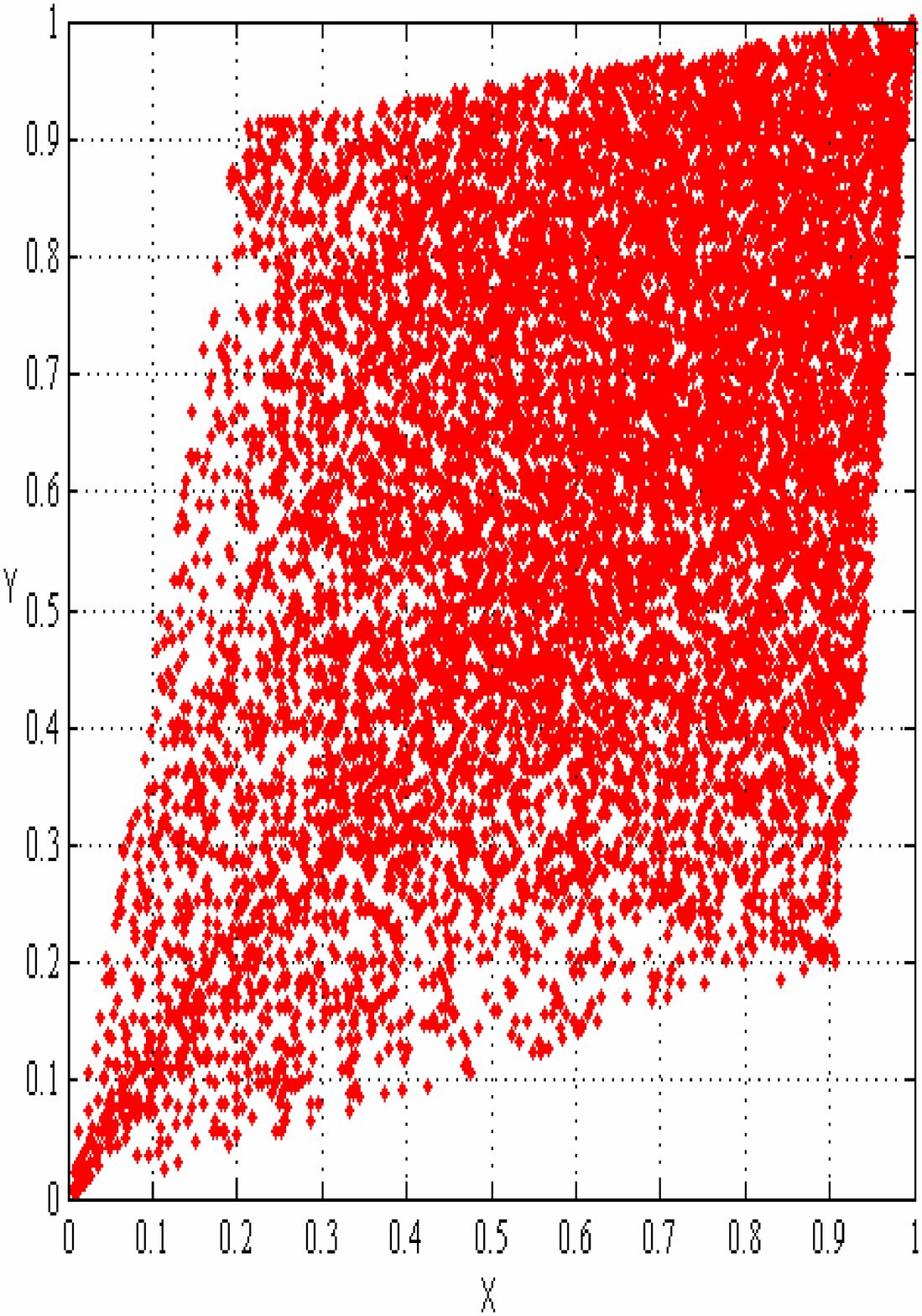}\includegraphics[bb=30mm 30mm 208mm 296mm, width=40.8mm, height=32.9mm, viewport=3mm 4mm 205mm 292mm]{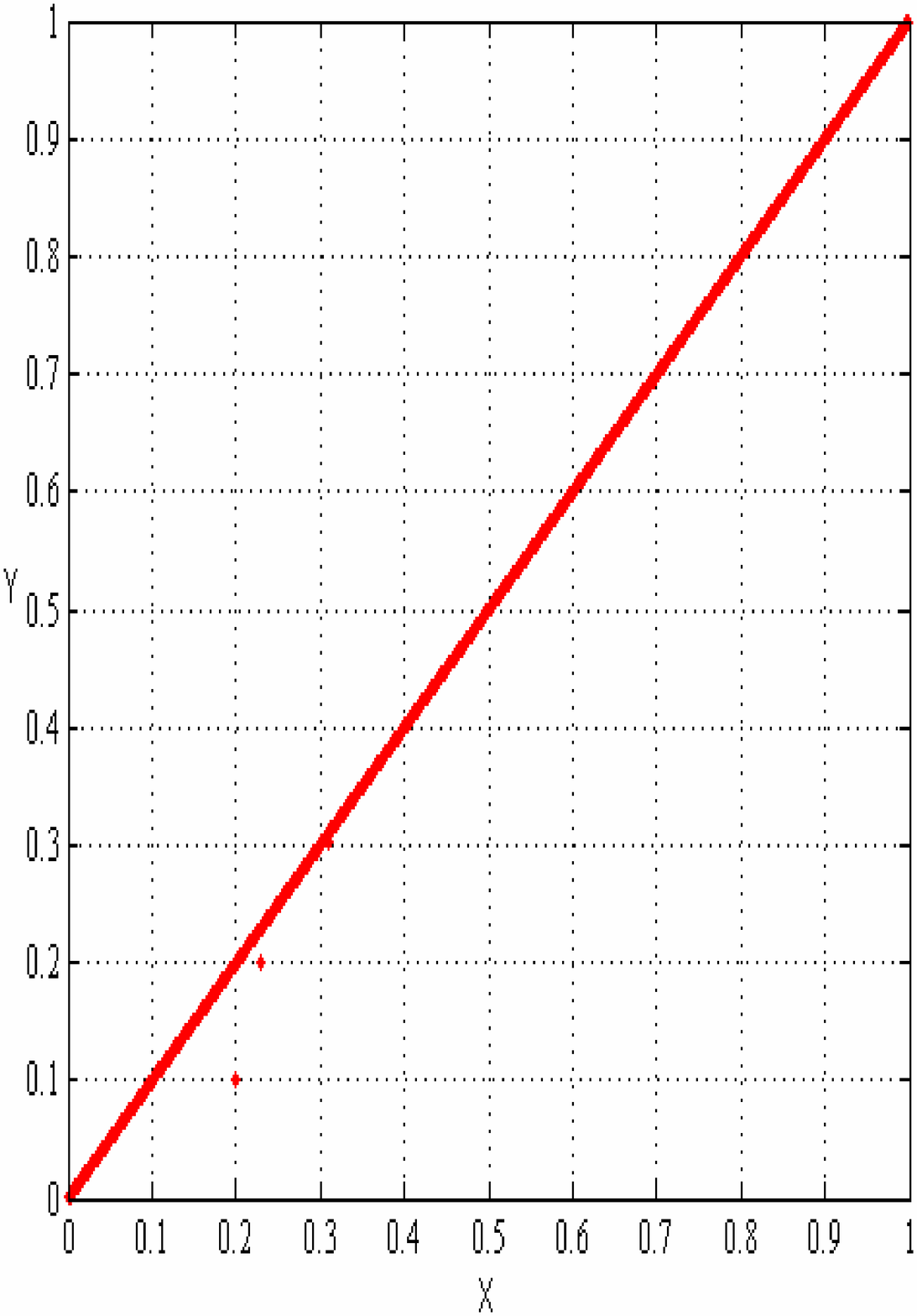}
      \caption{Chaotic oscillators coupled with different
      strengths. Left: \textit{e} = 0.9, i.e. outside
      of the contraction range. Right: \textit{e} = 0.4, i.e inside
      the predicted synchronization interval.}
      \label{figurelabel}
\end{figure}

\subsection{Coupled Map Lattice (CML)}

In [5], the authors proposed a more general model a coupling in
which the states takes into account the dynamics of any other
oscillator ``next to it'' equally weighted:

\begin{equation} \label{GrindEQ__15_} x^{i} _{n+1} =\left(1-\varepsilon \right)f\left(x^{i} _{n} \right)+\frac{\varepsilon }{2} \left(f\left(x^{i+1} _{n} \right)+f\left(x^{i-1} _{n} \right)\right) \end{equation}

As it can be seen, at the iteration \textit{n+1} the dynamics of the
\textit{i-th} oscillator depends on its previous state (weighted by
\textit{1-$\varepsilon$}) and the previous state of the ``nearest''
two oscillators, e.g. \textit{(i-1)-th} oscillator and the
\textit{(i+1)-th} oscillator, equally weighted by a factor of
\textit{$\varepsilon$ /2}. Let \textit{$x_{n+1}$= f(x) =1 --
a$x_{n}^{2}$} where \textit{a} is a positive constant and write the
system in the following form:

\begin{equation} \label{GrindEQ__16_} x^{i} _{n+1} =f\left(x^{i} _{n} \right)+\frac{\varepsilon }{2} \left(f\left(x^{i+1} _{n} \right)+f\left(x^{i-1} _{n} \right)-2f\left(x^{i} _{n} \right)\right) \end{equation}

Thus, for two coupled oscillators \textit{x} and \textit{y}, the
matrices \textbf{U} and \textbf{V} are vectors

\[\begin{array}{l} {\textbf{U}=\left[\begin{array}{cc} {1} & {1} \end{array}\right]} \\ {\textbf{V}=\left[\begin{array}{cc} {1} & {-1} \end{array}\right]} \end{array}\]

and the dynamics of the projection on the synchronization manifold
\textit{$x^{\bot}$ =} \textit{e} = \textit{x} -\textit{ y}  is given
by

\begin{equation} \label{GrindEQ__17_} e_{n+1} =x_{n+1} -y_{n+1} =\left(\frac{3\varepsilon }{2} -1\right)a\sigma e_{n}\end{equation}

where \textit{$\sigma$ = x+y},  and no approximations, e.g.
linearization about a certain point, have been made. Now, a virtual
auxiliary system can be constructed [17]

\begin{equation} \label{GrindEQ__18_} w_{n+1} =f\left(w_{n} \right)=\left(\frac{3\varepsilon }{2} -1\right)a\sigma w_{n}  \end{equation}

where both 0 and \textit{$e_{n}$} are particular solutions. This new
map shall be contracting, i.e.
$\partial$\textit{f}(\textit{$w_{n}$})/$\partial$\textit{$w_{n}$} is
\textit{u.n.d} , in any case for an identity metric, i.e. in ($7$)
the matrix \textit{M} = \textit{I }(identity matrix), and therefore
\textit{e} will tend exponentially to 0 if [7]:

\begin{equation} \label{GrindEQ__19_} \frac{2\left(a-\frac{1}{\sigma } \right)}{3a} <\varepsilon <\frac{2\left(a+\frac{1}{\sigma } \right)}{3a}  \end{equation}

Letting \textit{$\sigma$} to be variable, ($19$) describes a family
of hyperbolas in the \textit{a} - $\varepsilon$ plane. Now, it is
necessary to find a region in the \textit{a} - $\varepsilon$ plane
that guarantees contraction, i.e. a region where any pair
(\textit{a} - $\varepsilon$) will assure contraction of the system.
This can be easily done by finding the supremum of the areas defined
between the curves of the hyperbolas. In Fig. 2 some of the
hyperbolas are shown for \textit{a} = 1.7, where the map exhibits a
chaotic behaviour (\textit{$\sigma_{1}$ $<$ $\sigma_{2}$ $<$
$\sigma_{3}$}).

\begin{figure}[thpb]
      \centering
      \includegraphics[bb=0mm 0mm 208mm 296mm, width=64.7mm, height=51.7mm, viewport=3mm 4mm 205mm 292mm]{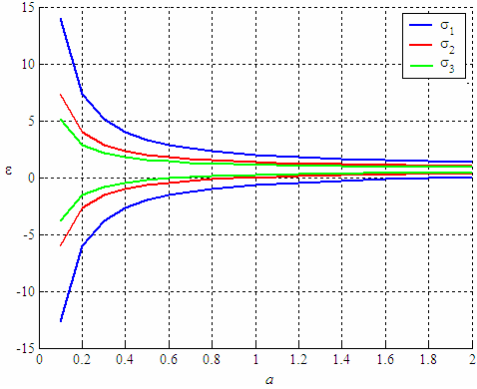}
      \caption{Family of hyperbolas defining contraction regions
      for the coupled system. The hyperbola with the ``narrowest'' area
      between the curves ($\sigma_{3}$) defines the sufficient contraction
      region for the error dynamics}
      \label{figurelabel}
\end{figure}

In this particular case, it is found that the supremum is given by
the area between the curves o the hyperbola with \textit{$\sigma$
=}2(2/\textit{$\sqrt{a}$}). Now, since\textit{ } the initial value
of \textit{x} is random number between 0 and 1, the absolute values
of \textit{x }and \textit{y} are always less than
(\textit{$\sqrt{2/a}$}) $\approx$ 1 (for values of a which guarantee
chaotic behaviour), it can be easily shown that:

\begin{equation} \label{GrindEQ__20_}  {-1<1-ax^{2} <1} \\\Rightarrow {-2<-ax^{2} <0} \end{equation}

Since both \textit{a} and \textit{$x^{2}$} are always positive, the
condition in ($20$) is true.

The values of \textit{a} which guarantee chaotic behaviour can be
found by computing the Lyapunov exponents of the map as \textit{a}
varies [16]. The first Lyapunov exponent for different values of
\textit{a} is shown in Fig. 3.

\begin{figure}[thpb]
      \centering
      \includegraphics[bb=0mm 0mm 208mm 296mm, width=63.7mm, height=50.8mm, viewport=3mm 4mm 205mm 292mm]{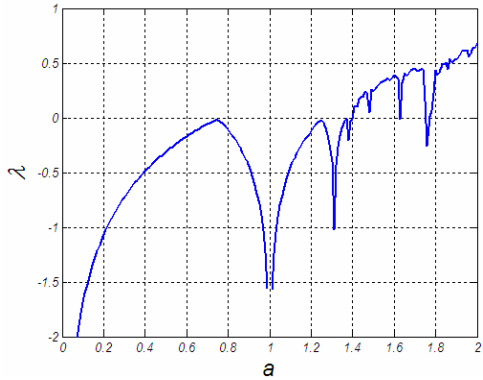}
      \caption{First Lyapunov exponent of the logistic map as
      \textit{a} varies}
      \label{figurelabel}
\end{figure}

It can be seen that the values close to 2 guarantee chaotic
behaviour. Hence, the limit, i.e. the narrowest hyperbola is given
by \textbf{$\sigma\approx2$} , since

\begin{equation} \label{GrindEQ__21_} \frac{2a-1}{3a} <\varepsilon <\frac{2a+1}{3a}  \end{equation}

It is worth noticing that this is the same result on would find
applying theorem $2$: writing the system of equations in the form:

\begin{equation} \label{GrindEQ__22_} x_{n+1} -\left(\frac{3\varepsilon }{2} -1\right)ax_{n} ^{2} =y_{n+1} -\left(\frac{3\varepsilon }{2} -1\right)ay_{n} ^{2}  \end{equation}

where, in order to assure synchronization of the states, the
function:

\begin{equation} \label{GrindEQ__23_} h\left(x\right)=\left(\frac{3\varepsilon }{2} -1\right)ax^{2}  \end{equation}

must be contracting in some metric. Using the identity metric the
same results as before are found.

Thus, the expected behaviour should be (looking for positive values
of \textit{$\varepsilon$ })  as presented in figure 4, where the
shaded area represents the guaranteed contraction region for the
coupled system (recall that each oscillator without coupling follows
its own nominal chaotic dynamics):

\begin{figure}[thpb]
      \centering
      \includegraphics[bb=0mm 0mm 208mm 296mm, width=50.3mm, height=41.9mm, viewport=3mm 4mm 205mm 292mm]{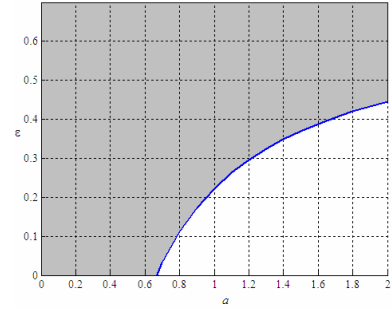}
      \caption{The shaded area is the guaranteed contraction
      region for $\sigma$= 2 and an identity metric}
      \label{figurelabel}
\end{figure}

Numerical experiments were performed: using a Sobolev distribution
(which gives a random but uniformly distributed cloud of points in a
determined region) in the \textit{a}-$\varepsilon$ plane, Fig. 4
shows the norm of the error vector for each couple
(\textit{a,$\varepsilon$}) after a transient time of 300 iterations.
We find indeed, as shown in figure 5, that the actual region of
guaranteed contraction for the coupled system resembles the one
predicted by the proposed method.

\begin{figure}[thpb]
      \centering
      \includegraphics[bb=0mm 0mm 208mm 296mm, width=52.9mm, height=40.6mm, viewport=3mm 4mm 205mm 292mm]{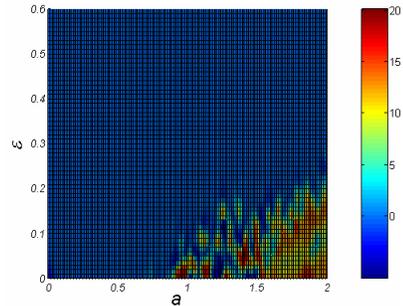}
      \caption{Actual contraction region for $\sigma$ = 2. The
      dark area represents a set of pairs (\textit{a,$\varepsilon$ })
      for which the norm of the error is zero after the transient}
      \label{figurelabel}
\end{figure}

\section{Simulations on the Coupled Map Lattice}

Using the previous results, the coupling strength can be tuned to
guaranteed synchronization, as shown in Fig. 6.

\begin{figure}[thpb]
      \centering
      \includegraphics[bb=0mm 0mm 208mm 296mm, width=77.6mm, height=60.2mm, viewport=3mm 4mm 205mm 292mm]{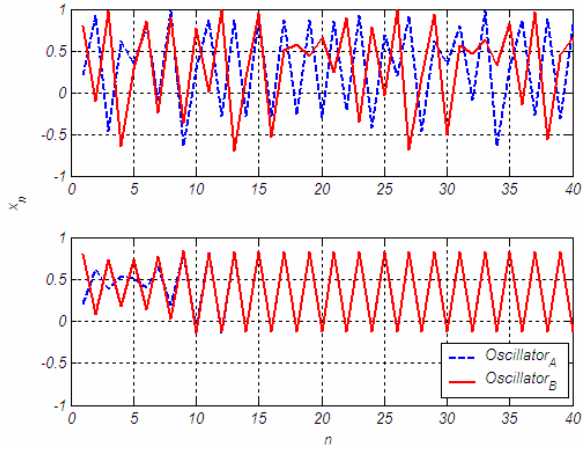}
      \caption{Two chaotic discrete oscillators. Above: non coupled. Below: properly coupled}
      \label{figurelabel}
\end{figure}

From these results, it is clear how, using a value for the coupling
strength that satisfy the conditions for the contracting error
dynamics in the synchronization manifold, the two systems will tend
exponentially to each other (equivalently, the error will tend
exponentially to the origin of the synchronization manifold).

\section{Adaptation and Emerging Behaviour Via Sensing}

Let's modify the coupled map lattice considered above such as the
system receives information of the environment via sensing. The
signal coming from the sensors can be any kind of signal (rotation
angle, voltage, etc) as long as it can be related somehow to the
state variable \textit{x}. Let's consider for example the following
dynamics as proposed in [5]:

\begin{scriptsize}
\begin{equation}\label{GrindEQ__24_} x^{i} _{n+1}=f\left(x^{i} _{n} +\varepsilon _{1} \left(\bar{s}_{n} -s^{i} _{n} \right)+\varepsilon _{2} \left(\frac{s^{i+1} _{n} +s^{i-1} _{n} }{2} -s^{i} _{n} \right)\right) \end{equation}
\end{scriptsize}

where \textit{$s^{i}_{n}$} denotes the sensor reading at the
iteration \textit{n} and  $\bar{s}_{n} $  is the mean value of the
sensors at the iteration \textit{n}. The meaning of this type of
coupling is as follows: each oscillator follows its own the chaotic
dynamics, but this is ``adjusted and updated'' to reduce the global
(via \textit{$\varepsilon_{1}$}) and local (via
\textit{$\varepsilon_{2}$}) difference in the sensors [5]. Fig. 7
shows the model schematically.

\begin{figure}[thpb]
      \centering
      \includegraphics[bb=0mm 0mm 208mm 296mm, width=75.6mm, height=53.5mm, viewport=3mm 4mm 205mm 292mm]{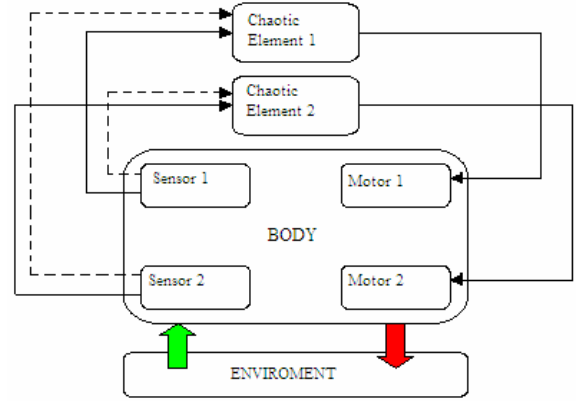}
      \caption{Outline of the model. This model is inspired on
      the one proposed in [5]}
      \label{figurelabel}
\end{figure}

Let's suppose that each oscillator represents the dynamics of a
motor and that each motor is driving a leg of a simple two--legged
robot. Thus, each leg has an actuator following the proposed
dynamics coupled in order to achieve synchronization. Let the
variable \textit{x} be the command to the actuator (torque $\tau$)
and the signal sensed (s) be the angle $\theta$. This to variables
are related as follows in Fig. 8:

\begin{figure}[thpb]
      \centering
      \includegraphics[bb=0mm 0mm 208mm 296mm, width=37.7mm, height=46.0mm, viewport=3mm 4mm 205mm 292mm]{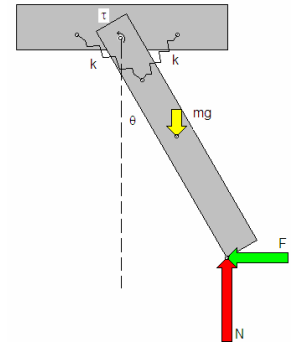}
      \caption{Model used for the two--legged robot simulation}
      \label{figurelabel}
\end{figure}

Following the method as done previously in (18), the error dynamics
is given by:
\begin{equation}\label{GrindEQ__25_} e_{n+1} =-a\sigma e_{n} +2a\left(\frac{\varepsilon _{1} }{2} +\varepsilon _{2} \right)\sigma (s_{n} ^{(A)} -s_{n} ^{(B)} ) \end{equation}

For this is particular case, the angle \textit{s} can be related
directly to the torque \textit{x} in a linear fashion, e.g. in
($25$)
\\\textit{s} = \textit{mx} + \textit{b} (where the constants
\textit{m} and \textit{b} are given by the sensor characteristics).
Hence once again, a virtual auxiliary system can be construct

\[y_{n+1} =f\left(y_{n} \right)=-a\sigma y_{n} +2a\left(\frac{\varepsilon _{1} }{2} +\varepsilon _{2} \right)\sigma (s_{n} ^{(A)} -s_{n} ^{(B)} )\]

where 0 and \textit{$e_{n}$} are particular solutions. Using an
identity metric \textit{M} = \textit{I}, the system shall be
contracting, i.e. the error \textit{e} will converge exponentially
to 0, for:

\begin{equation} \label{GrindEQ__26_} \begin{array}{l} {\frac{\psi (a-\frac{1}{\sigma } )}{2a} <\frac{\varepsilon _{1} }{2} +\varepsilon _{2} <\frac{\psi (a+\frac{1}{\sigma } )}{2a} {\rm \; \; \; \; \; \; \; }}
\bigskip
\\ {where{\rm \; \; \; \; \; \; \; \; }\psi =\left(K^{*} -Nl+mg\frac{l}{2} \right)} \end{array}
\end{equation}

and $K^{*}$ is the equivalent stiffness of the two springs. Fig. 9
shows two coupled oscillators starting at random initial conditions.
The complete synchronization is achieved by tuning the coupling
strength to be inside the predicted synchronization range.

\begin{figure}[thpb]
      \centering
      \includegraphics[bb=0mm 0mm 208mm 296mm, width=40mm, height=41.8mm, viewport=3mm 4mm 205mm 292mm]{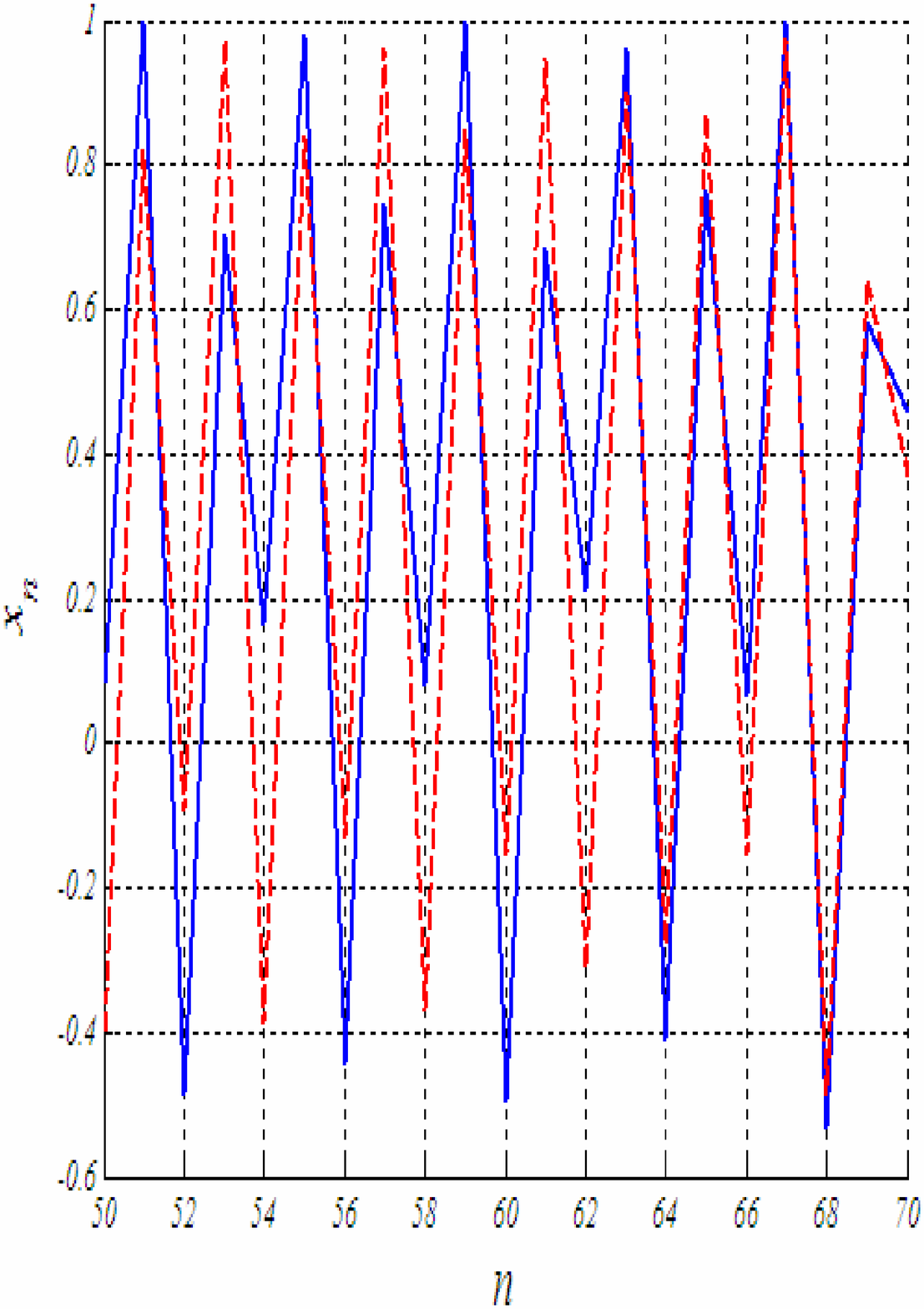}\includegraphics[bb=0mm 0mm 208mm 296mm, width=40mm, height=41.9mm, viewport=3mm 4mm 205mm 292mm]{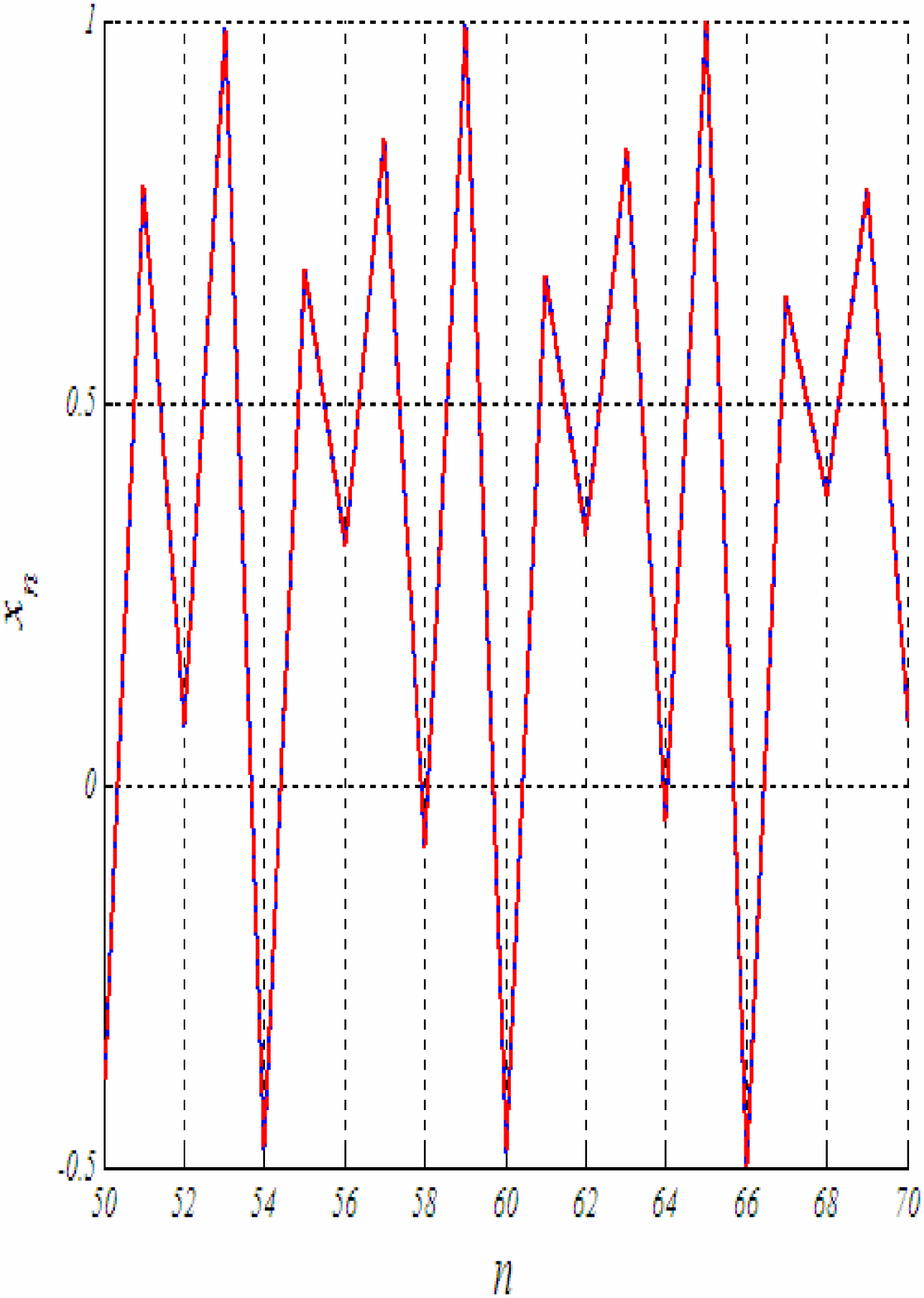}
      \caption{Proper selection of the coupling strength assures
      complete synchronization. Left: coupling strength outside the
      contraction range. Right: coupling strength inside the contraction
      range}
      \label{figurelabel}
\end{figure}

\section{Why Bother with Chaotic Systems?}

At this point it seems natural to ask why it is worth working in a
chaotic regime, since similar results, specifically synchronization,
can be achieved with non chaotic oscillators. Apparently the key to
answer this questions is adaptability. The following example shows
the advantages of a chaotic regime. Assume the two--legged robot
under three different conditions: first assume that the legs are
chaotic but not coupled; second, the legs are coupled but not
chaotic; and finally the legs are coupled working in the chaotic
regime. As an external condition assume that the left leg senses a
decrease of 25\% in the friction coefficient after y = 1 m. Figure
10 shows the results obtained by simulation.

\begin{figure}[thpb]
      \centering
      \includegraphics[bb=0mm 0mm 208mm 296mm, width=74.7mm, height=59.4mm, viewport=3mm 4mm 205mm 292mm]{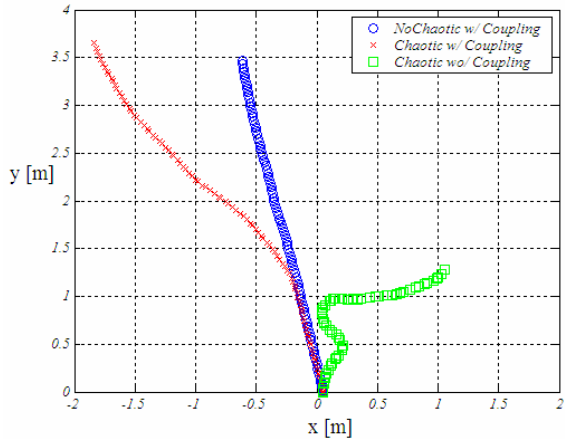}
      \caption{Trajectory followed by the robot in the x--y plane}
      \label{figurelabel}
\end{figure}

Apparently, the condition coupled--chaotic is more sensitive to
external changes in the environment, which can be seen as an
emerging adaptive property of the system.

Another simulation under the condition coupled--chaotic was done in
which after some fixed traveling distance one of the legs ``feels''
a change in the surface, e.g. in this case the sensor detects a
change in the angle due to a decrease in the friction coefficient
between the leg and the surface. The results suggest that the change
in the behaviour depends on the magnitude of the change in the
environment.

\begin{figure}[thpb]
      \centering
      \includegraphics[bb=0mm 0mm 208mm 296mm, width=42mm, height=39.1mm, viewport=3mm 4mm 205mm 292mm]{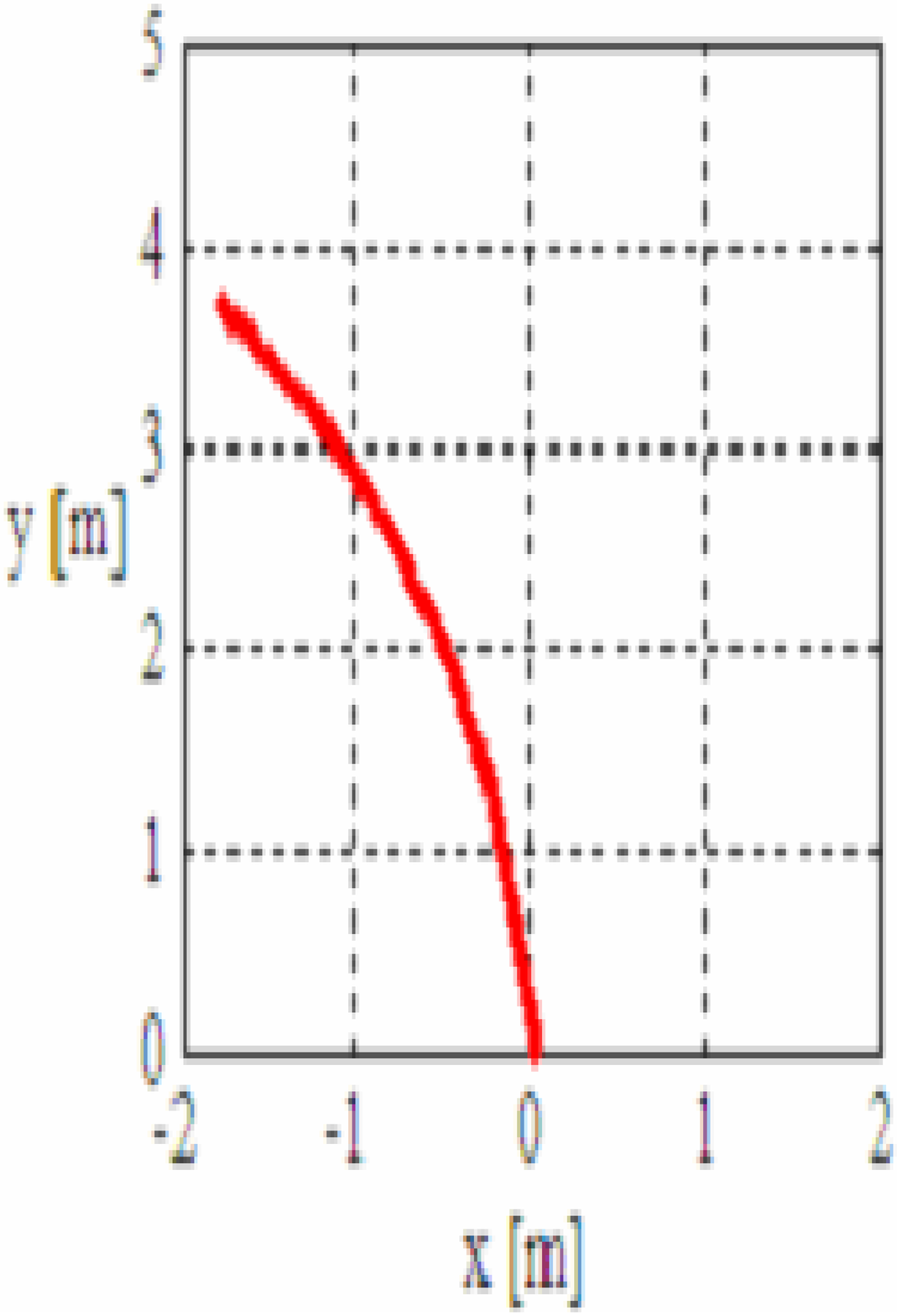}\includegraphics[bb=0mm 0mm 208mm 296mm, width=42mm, height=41.7mm, viewport=3mm 4mm 205mm 292mm]{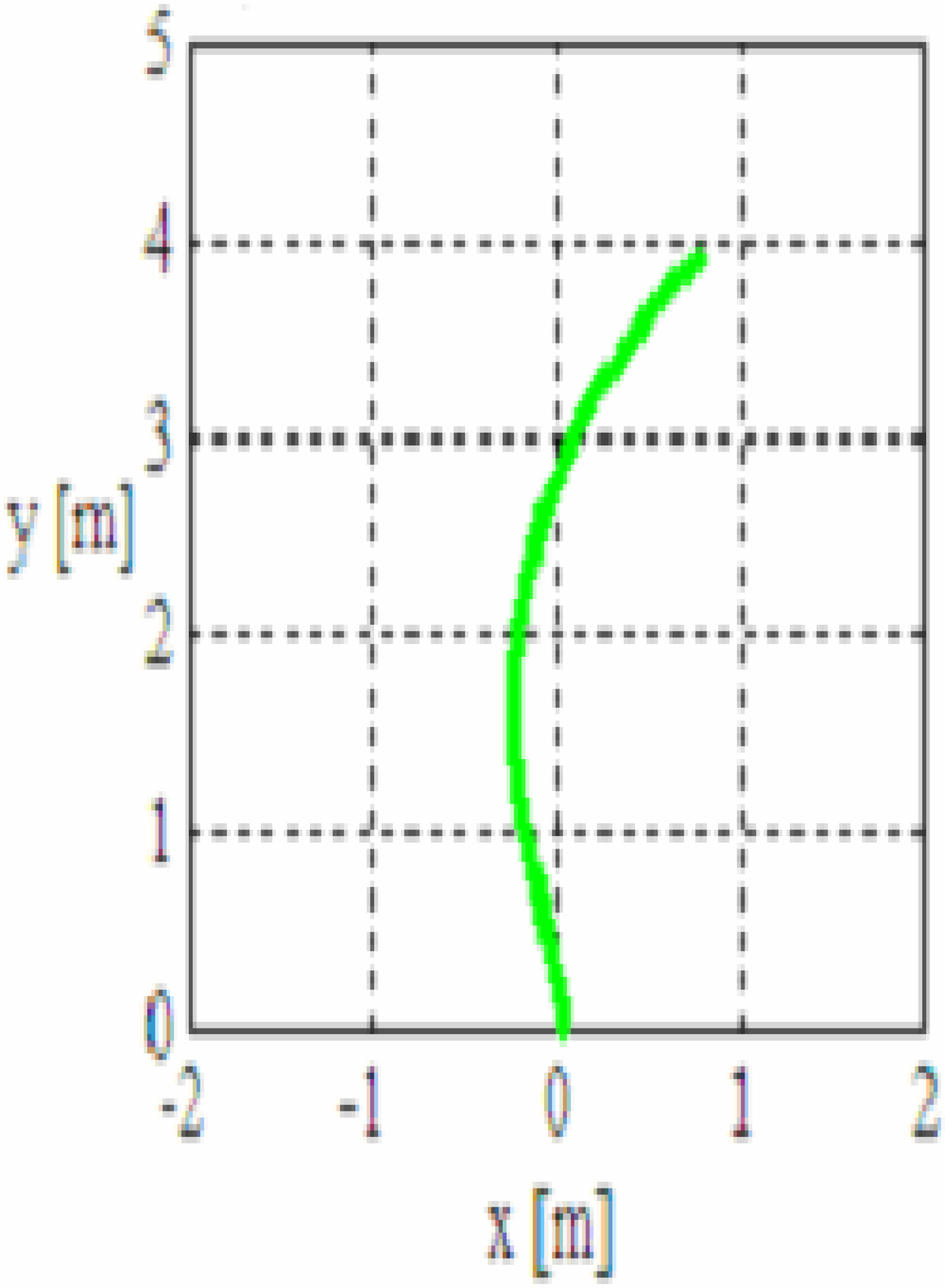}
      \caption{Friction coefficient between leg and surface
      decrease after        y = 1. Left: decrease of 25\% in the friction coefficient in the left leg. Right:
      decrease of 40\% in the friction coefficient in the right leg}
      \label{figurelabel}
\end{figure}

These results suggest that the idea of adaptation to changes in the
environment is intrinsic to the system dynamics. In the sense of
adaptation and evolution [$5$] and perhaps abusing of the
terminology, chaotic oscillations may be regarded as ``mutants''
from a nominal (non chaotic) oscillation pattern. Such deviations
from the smooth nominal behaviour give the system the capacity of
response faster to external inputs. Furthermore, due to the nature
of chaotic oscillations, e.g. strange attractors where the
trajectories are random--like yet confined in space [15], they
appear to be a good compromise between a ``mutation'' sufficiently
strong to make the system adapt to a new external condition but mild
enough to maintain information from the previous behaviour.

Finally, note that the analysis does not require any goal direction,
because the system, due to its simple configuration, tends to ``walk
straight ahead''. This feature makes unnecessary the implementation
of any mean of command, e.g. algorithms deciding to freeze a leg at
a certain moment. The system simply moves forward, responding
quickly to changes in the environment and, more interesting, this
response is proportional to the magnitude of the change.

But the reason why the chaotic regime exhibits this adaptive
behaviour remains an open question. In order to understand this
phenomenon let's take the following example: assume that each
movement of the legs takes place in 1 second (arbitrary time
interval where one oscillation takes place). It is obvious that the
power spectral density of the signal (PSD) will have, in any case, a
peak at 0.5 Hz (it takes 2 seconds for the leg to complete a cycle
forward-backward).

\begin{figure}[thpb]
      \centering
      \includegraphics[bb=0mm 0mm 208mm 296mm, width=66.9mm, height=52.9mm, viewport=3mm 4mm 205mm 292mm]{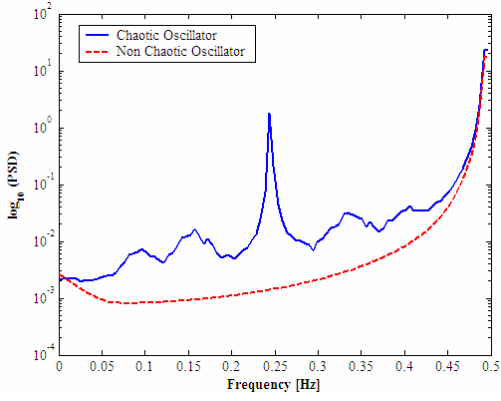}
      \caption{Power spectral density of a chaotic and a
      non-chaotic oscillator}
      \label{figurelabel}
\end{figure}

Fig. 12 shows the mean PSD of 128 runs starting at random initial
conditions: the mean of 128 non chaotic signals, i.e. with a value
of\textit{ a }that makes the first Lyapunov exponent negative, e.g.
\textit{a} = 1.1 (dashed line) and the mean of 128 chaotic signals,
i.e. \textit{a} = 1.7 (solid line). Both signals are subjected to a
equivalent coupling strength

It can be seen how the chaotic signal carries energy at some other
different frequencies than the trivial one. In particular, it can be
seen that the period-doubling phenomenon occurs (peak at half the
main frequency). Due to its chaotic nature, this particular feature
allows the system to have richer movement patterns in changing
environments (because different frequencies are being excited by the
same signal).

For example, consider the case of a small vehicle where each wheel
is driven by a chaotic oscillator and these oscillators are coupled
as before. As any road profile irregularity can be describe as an
ergodic random process in space [2], a vehicle traveling at constant
speed \textit{V} (in average) will sense an input which is an
ergodic random process in time. As the energy of the signal in
chaotic regime is amplified at more particular frequencies, the
natural response of the system, e.g. the legged robot or the wheeled
vehicle, will show richer outputs. This could be seen as an
advantage when dealing with challenging and changing environments.

\section{Networks of Symmetric Coupled Oscillators}
Up to this point, this work has focused on two coupled oscillators.
However, the extension to more oscillators in a symmetric network is
straight forward. Indeed, as it will be shown in this section, the
synchronization analysis of this particular type of networks can be
done as if it were composed by two oscillators.

Let's start by saying that matrices \textbf{U} and \textbf{V} in
($2$) and ($3$) can always be defined for \textit{n} identical
oscillators with scalar dynamics and symmetric coupling, leading to
a synchronization of the states of any two adjacent oscillators.

Now, suppose \textit{n} identical oscillators, coupled in such way
that the each oscillator updates its own dynamics by sensing the
dynamics of every other oscillator in the network. Furthermore,
suppose that the magnitude of the coupling strength is equal for any
pair of oscillators in the network. The structure of a network with
verifying such conditions is known as ``all--to--all'' symmetry
structure [17]. As a general case, consider the Coupled Map Lattice:

\begin{footnotesize}
\begin{equation} \label{GrindEQ__27_} \left[\begin{array}{c} {x^{\left(1\right)} _{t+1} } \\ {x^{\left(2\right)} _{t+1} } \\ {\vdots } \\ {x^{\left(n\right)} _{t+1} } \end{array}\right]=\left[\begin{array}{cccc} {1-\varepsilon } & {g\left(\varepsilon \right)} & {\cdots } & {g\left(\varepsilon \right)} \\ {g\left(\varepsilon \right)} & {1-\varepsilon } & {\cdots } & {g\left(\varepsilon \right)} \\ {\vdots } & {\vdots } & {\ddots } & {\vdots } \\ {g\left(\varepsilon \right)} & {g\left(\varepsilon \right)} & {\cdots } & {1-\varepsilon } \end{array}\right]\left[\begin{array}{c} {f\left(x^{\left(1\right)} _{t} \right)} \\ {f\left(x^{\left(2\right)} _{t} \right)} \\ {\vdots } \\ {f\left(x^{\left(n\right)} _{t} \right)} \end{array}\right] \end{equation}
\end{footnotesize}

where \textit{f}(\textit{$x_{t}^{(k)}$}) is any function
representing a discrete dynamical system (for \textit{k} = 1,
2,\dots , \textit{n}) and \textit{g}(\textit{$\varepsilon$}) is a
general function of the coupling strength. Using
(\ref{GrindEQ__4_}), it is possible to find the projection of the
system on the transverse manifold, leading to a (\textit{n} -- 1)
dynamics of the form

\begin{equation}\label{GrindEQ__28_}e_{n+1}=\Upsilon e_{n}\end{equation}

where $\Upsilon$ is a diagonal matrix where the diagonal elements
are functions of the form
\textit{$\upsilon_{jj}$}(\textit{$\varepsilon$},\textit{
$x_{t}^{(j)}$} ,\textit{ $x_{t}^{(j+1)})$} for \textit{j} = 1,
2,\dots , \textit{n}--1. Thus, each element of the diagonal is a
function that can be expressed for any pair of oscillators in a
``two-by-two'' fashion. If \textit{$\upsilon_{jj}$} is \textit{upper
bounded}, then the coupling strength \textit{$\varepsilon$} can
\textit{always} be tuned to guaranteed uniformly negative
definiteness of the diagonal matrix $\Upsilon$ [$17$]. It is worth
noticing that in the case of chaotic oscillators the condition on
the upper bounded of the function is guaranteed by the strange
attractor behaviour [$16$], as agreed in section VIII.

This results leads to a convenient simplification in the analysis of
any symmetric network of oscillators, since it is necessary to study
one single function \textit{g} to guarantee complete
synchronization. Furthermore, it can be stated that the
synchronization analysis of any network of oscillators with an
``all--to--all'' symmetry structure can be made studying the
uniformly negative definiteness of a single \textit{scalar}
function. For example, for three discrete oscillators \textit{x},
\textit{y} and \textit{z}:

\begin{small}
\begin{equation} \label{GrindEQ__29_} \begin{array}{l} {x_{n+1} =f\left(x\right)=1-ax_{n} ^{2} } \\ {y_{n+1} =f\left(y\right)=1-ay_{n} ^{2} } \\ {z_{n+1} =f\left(z\right)=1-az_{n} ^{2} } \end{array} \end{equation}
\end{small}

working in chaotic regime, e.g. \textit{a} = 1.7, coupled by a
linear operator in the form

\begin{small}
\begin{equation} \label{GrindEQ__30_} \left[\begin{array}{c} {x_{t+1} } \\ {\begin{array}{l} {y_{t+1} } \\ {z_{t+1} } \end{array}} \end{array}\right]=\left[\begin{array}{ccc} {1-\varepsilon } & {\frac{\varepsilon }{2} } & {\frac{\varepsilon }{2} } \\ {\frac{\varepsilon }{2} } & {1-\varepsilon } & {\frac{\varepsilon }{2} } \\ {\frac{\varepsilon }{2} } & {\frac{\varepsilon }{2} } & {1-\varepsilon } \end{array}\right]\left[\begin{array}{c} {f\left(x_{t} \right)} \\ {\begin{array}{l} {f\left(y_{t} \right)} \\ {f\left(z_{t} \right)} \end{array}} \end{array}\right] \end{equation}
\end{small}

can be analyzed as previously, e.g. \textit{$x_{1}^{\bot}$} =
\textit{$e^{(1)}$} = \textit{x} -- \textit{y} and
\textit{$x_{2}^{\bot}$} =\textit{$e^{(2)}$} = \textit{y} --
\textit{z} leading to

\begin{footnotesize}
\begin{equation} \label{GrindEQ__31_} \left[\begin{array}{l} {e^{\left(1\right)} _{n+1} } \\ {e^{\left(2\right)} _{n+1} } \end{array}\right]=\left[\begin{array}{cc} {\left(\varepsilon -1\right)a\sigma _{1} } & {0} \\ {0} & {\left(\varepsilon -1\right)a\sigma _{2} } \end{array}\right]\left[\begin{array}{l} {e^{\left(1\right)} _{n} } \\ {e^{\left(2\right)} _{n} } \end{array}\right]=Fe_{n}  \end{equation}
\end{footnotesize}

where \textit{$\sigma_{1}$} = (\textit{x} + \textit{y}) and
\textit{$\sigma_{2}$} = (\textit{y} + \textit{z}). Once again, a
virtual auxiliary system in the form

\begin{tiny}
\begin{equation} \label{GrindEQ__32_} \left[\begin{array}{l} {p_{n+1} } \\ {q_{n+1} } \end{array}\right]=\left[\begin{array}{cc} {\left(\varepsilon -1\right)a\sigma _{1} } & {0} \\ {0} & {\left(\varepsilon -1\right)a\sigma _{2} } \end{array}\right]\left[\begin{array}{l} {p_{n} } \\ {q_{n} } \end{array}\right]=F\left[\begin{array}{l} {p_{n} } \\ {q_{n} } \end{array}\right] \end{equation}
\end{tiny}

for which the vectors \textbf{0} and \textbf{$e_{n}$} are particular
solutions. Thus, the states will converge exponentially to each
other for a coupling strength \textit{$\varepsilon$} that guarantees
uniformly negative definiteness of the matrix \textbf{F}, i.e. the
largest singular value of \textbf{F} remains smaller than 1
uniformly [3]. This is easily done (due to the diagonal form of the
matrix \textbf{F}) by tuning the parameter \textit{$\varepsilon$} in
the function

\begin{equation} \label{GrindEQ__33_} \gamma \left(\varepsilon \right)=\left(\varepsilon -1\right)\sigma _{k} a \end{equation}

so that \textit{$\gamma$} is uniformly negative definite. The upper
bounded value of \textit{$\sigma_{k}$} is easily checked from the
previous examples (\textit{$\sigma_{k}\approx2$}).

Finally, for more general systems, the system of equations in
(\ref{GrindEQ__29_}) can always be expressed in such way that the
following theorem from classical contraction analysis can be
directly applied [17].

\bigskip

\textit{\underbar{Theorem 3}}

Consider \textit{q} coupled systems. IF a contracting function
\textit{h}(\textit{$x_{i}$}) exists such that

\[x^{\left(1\right)} _{n+1} -h\left(x^{\left(1\right)} _{n} \right)=x^{\left(2\right)} _{n+1} -h\left(x^{\left(2\right)} _{n} \right)=...=x^{\left(q\right)} _{n+1} -h\left(x^{\left(q\right)} _{n} \right)\]

THEN all the systems synchronize exponentially, regardless of the
initial conditions.

\bigskip

In the case of the last example, the system dynamics can be
expressed as

\begin{footnotesize}
\begin{equation} \label{GrindEQ__35_} x_{t+1} -\left(\varepsilon -1\right)ax^{2} _{t} =y_{t+1} -\left(\varepsilon -1\right)ay^{2} _{t} =z_{t+1} -\left(\varepsilon -1\right)az^{2} _{t}  \end{equation}
\end{footnotesize}

for which the analysis of the function \textit{h}(\textit{$x_{t}$})
= (\textit{$\varepsilon$ }-- 1)\textit{$ax_{t}^{2}$} and the tuning
of \textit{$\varepsilon$} to achieve uniformly negative definiteness
of \textit{h}, leads to the same result as above.

\section{CONCLUSIONS AND FUTURE WORKS}

\subsection{Conclusions}

This paper has shown the feasibility of studying synchronized
discrete chaotic oscillators, by means of a differential analysis of
the nonlinear system. Different kinds of networks and coupling
operators can be analyzed globally by simple method. Further
investigations on the emerging adaptation properties of such systems
are necessary for practical implementations.

\subsection{Future Works}

Future work will be focus on the extension to systems with more
degrees of freedom. Some of the work done on synchronization is
intended to be implemented in a set of Lego
Mindstorm$^{\rlap{$\bigcirc$}\scriptstyle\,\,\rm R\;}$ . A small
two-wheel vehicle was built using the set of
Lego$^{\rlap{$\bigcirc$}\scriptstyle\,\,\rm R\;}$  blocks and the
input signal to the two motors driving the wheels is given by
($26$). The algorithm is intended to be implemented using the angle
of rotation as the sensed signal, i.e in accordance with the model
of the two-legged robot presented in simulations.

\section{ACKNOWLEDGMENTS}

Juan C. Botero thanks Prof. G. Mastinu and Prof. M. Gobbi for their
support and the NSL staff at MIT for their constant help.


\begin{thebibliography}{99}

\bibitem{c1}
P. Ashwin. ``Synchronization from Chaos''. \textit{Nature} 422,
2003, pp. 384 - 385.

\bibitem{c2}
M. Gobbi and G. Mastinu. ``Analytical Description and Optimisation
of the Dynamic Behaviour of Passively Suspended Road Vehicles''.
\textit{Journal of Sound and Vibration}, 245 (3), 2001, pp. 457 -
481.

\bibitem{c3}
E.M. Izhikevich. \textit{Dynamical Systems in Neuroscience: The
Geometry of Excitability and Bursting.} The MIT Press. Cambridge.
2007.

\bibitem{c4}
S. Kauffman. ``Requirements for evolvability in complex systems:
Orderly dynamics and frozen components''.\textit{ Physica D} 42,
1990, pp. 135 - 152.

\bibitem{c5}
Y. Kuniyoshi and S.Suzuki. ``Dynamic Emergence and Adaptation
Behavior Through Embodiment as Coupled Chaotic Field''.
\textit{Proceedings of 2004 IEEE/RSJ International Conference on
Intelligent Robots and Systems}. Sendai, Japan. 2004.

\bibitem{c6}
S.A.Levin. ``Complex Adaptive Systems: Exploring the known, the
unknown and the unknowable''. \textit{Bulletin of the American
Mathematical Society} 40 (1), 2002. pp. 3 - 19.

\bibitem{c7}
W. Lohmiller and J.J.E. Slotine. ``On Contraction Analysis for
Nonlinear Systems'' \textit{Automatica} 34 (6), 1998.

\bibitem{c8}
M. Mitchell, J. Hraber and P. Crutchfield. ``Revisiting the Edge of
Chaos: Evolving Cellular Automata to Perform Computations''.
\textit{Santa Fe Institute}. 1993. Santa Fe Institute Working Paper
\# 93--03--014.

\bibitem{c9}
H. Nijmeijer. ``A dynamical control view on synchronization''
\textit{Physica D} 154, 2001. pp. 219 - 228.

\bibitem{c10}
L.M. Pecora, T.L. Carroll, G. Johnson, D.J. Mar and J.F Heagy.
``Fundamentals of Synchronization in Chaotic Systems, Concepts and
Applcations''. \textit{Chaos} 7 (4), 1997.

\bibitem{c10}
L.M. Pecora and T.L. Carroll. ``Driving Systems with Chaotic
Signals''. \textit{Physical Review A} 44 (4), 1991.

\bibitem{c11}
A.S. Pikovsky, M.G. Rosenblum and J. Kurths, ed.
\textit{Synchronization. A Universal Concept in Nonlinear Sciences}.
Cambridge University Press. 2001.

\bibitem{c12}
Q. Pham and J.J.E. Slotine. ``Stable Concurrent Synchronization in
Dynamics System Networks''. MIT-NSL Report, 2005.

\bibitem{c13}
J.J.E. Slotine. ``Modular Stability Tools for Distributed
Computation and Control''.\textit{ Int. J. Adaptive Control and
Signal Processing}, 17(6), 2003.

\bibitem{c14}
E. Solak. ``A reduced-order observer for the synchronization of
Lorenz systems''. \textit{Physycs Letters A} 325, 2004. pp. 276 -
278.

\bibitem{c15}
S. Strogatz. \textit{Nonlinear Dynamics and Chaos: with Applications
to Physics, Biology, Chemistry and Engineering}. Addison Wesley
Publ. 1994.

\bibitem{c16}
W. Wang and J.J.E. Slotine. ``On Partial Contraction Analysis for
Coupled Nonlinear Oscillators''. \textit{Biological Cybernetics,} 92
(1), 2004.

\end{thebibliography}
\end{document}